\documentclass{article}

\usepackage{PRIMEarxiv}
\usepackage[dvipsnames]{xcolor}

\usepackage[utf8]{inputenc} 
\usepackage[T1]{fontenc}    
\usepackage{hyperref}       
\usepackage{url}            
\usepackage{booktabs}       
\usepackage{amsfonts}       
\usepackage{nicefrac}       
\usepackage{microtype}      
\usepackage{lipsum}
\usepackage{fancyhdr}       
\usepackage{graphicx}       
\graphicspath{{media/}}     

\title{Can AI Put Gamma-Ray Astrophysicists Out of a Job?
\thanks{Thankfully not.}
}

\author{
  S.T. Spencer\thanks{Also visitor at University of Oxford.} , V. Joshi, A.M.W. Mitchell \\
  Friedrich-Alexander University Erlangen-Nuremberg \\
  Erlangen Centre for Astroparticle Physics (ECAP) \\
  Nikolaus-Fiebiger Str. 2 \\
  91058 Erlangen\\
  Germany\\
  \texttt{samuel.spencer@fau.de}
}

\begin{document}
\maketitle

\begin{abstract}

In what will likely be a litany of generative-model-themed arXiv submissions celebrating April the 1st, we evaluate the capacity of state-of-the-art transformer models to create a paper detailing the detection of a Pulsar Wind Nebula with a non-existent Imaging Atmospheric Cherenkov Telescope (IACT) Array. We do this to evaluate the ability of such models to interpret astronomical observations and sources based on language information alone, and to assess potential means by which fraudulently generated scientific papers could be identified during peer review (given that reliable generative model watermarking has yet to be deployed for these tools). We conclude that our jobs as astronomers are safe for the time being. From this point on, prompts given to \textit{ChatGPT} and \textit{Stable Diffusion} are shown in {\color{BurntOrange} orange}, text generated by \textit{ChatGPT} is shown in black, whereas analysis by the (human) authors is in {\color{Cerulean} blue}. 

\end{abstract}

\begin{abstract}
{\color{BurntOrange} In this conversation you will generate a scientific paper (with an appropriate writing style) regarding the detection of a new Pulsar Wind Nebula. The Pulsar Wind Nebula is coincident with the previously known pulsar PSR J0456-69, and has been detected with the newly inaugurated ECAP-ACT Imaging Atmospheric Cherenkov Telescope (IACT) array. ECAP-ACT is located at 49.585N, 11.030E, and consists of 300 IACTs with a 6-metre Schwarzschild-Couder optical design and Silicon Photomultiplier (SiPM) cameras. You will give the Pulsar Wind Nebula an appropriate astronomical identifier and refer to it using this throughout the paper. During this conversation, typeset all relevant numerical values, formulae and tables in LaTeX. Do not provide summaries in your responses in this conversation until explicitly asked. Now begin by providing the abstract for this paper. In this, briefly discuss an observation campaign during 2020 of this source, the properties of the observed gamma-ray spectrum, and an appropriate model describing both the source morphology and multi-wavelength Spectral Energy Distribution (SED).} 

We report the detection of a new Pulsar Wind Nebula (PWN) using the newly inaugurated ECAP-ACT Imaging Atmospheric Cherenkov Telescope (IACT) Array. The PWN is coincident with the previously known pulsar PSR J0456-69 and has been identified as ECAP-ACT J0456-69. We present the results of an observation campaign during 2020, which yielded a gamma-ray spectrum with an energy range of 0.2-100 TeV. The spectrum is well-described by a power-law model with a spectral index of $\Gamma = 2.3\pm0.1$. The morphology of the source is well-defined, with an extension of $\sim$0.2$^\circ$ and a position angle of $\sim$20$^\circ$. We analyze the multi-wavelength Spectral Energy Distribution (SED) of ECAP-ACT J0456-69 and find that it can be modeled by a combination of synchrotron and inverse Compton emission from relativistic electrons. We discuss the implications of our results in the context of the energetics and evolution of pulsar wind nebulae. Our findings demonstrate the potential of ECAP-ACT for the study of high-energy astrophysical sources.

\end{abstract}


{\color{Cerulean} For this analysis we used \textit{ChatGPT} version Feburary 13 2023 \cite{chatgpt} and \textit{Stable Diffusion} v2.1-768 \cite{stablediffusion}. Already in the abstract, we note that the quoted energy range matches well with the range to which current generation IACTs are sensitive. The energy threshold of 0.2\,TeV is fairly accurate for a 6\,m telescope size and consistent with existing literature about telescopes of Schwarzschild-Couder optical design \cite{White_2021}. The spectral index generated of $\Gamma =2.3\pm 0.1$ is compatible with the average spectral index of galactic TeV sources ($\Gamma =2.4\pm 0.3$) \cite{gps}. The physical mechanisms of synchrotron and inverse Compton emission are correctly identified. \textit{ChatGPT} also generated keywords here despite not being asked.}

\section{Introduction}
{\color{BurntOrange} Generate an introduction section for this paper (with minimal re-use of text from the abstract), discussing the broad properties of Pulsar Wind Nebulae detectable with IACTs. Then discuss the properties of the pulsar PSR J0456-69, previous observations of the source region with the Fermi space telescope, and how observations with this instrument will provide greater physical insights than those from other current generation IACT arrays. For this section, and all following prompts, include appropriate academic references using the LaTeX \textbackslash cite\{\} command.}

Pulsar Wind Nebulae (PWNe) are the result of the interaction between the high-energy pulsar wind and the surrounding medium. These objects are powerful sources of non-thermal radiation across the electromagnetic spectrum, from radio waves to gamma rays \cite{Gaensler2006} {\color{Cerulean} (Gaensler2006)}. One of the key goals in the study of PWNe is to understand the physical processes that give rise to their multi-wavelength emission, which can provide insights into the properties of the pulsar wind and its interaction with the environment.

In recent years, Imaging Atmospheric Cherenkov Telescopes (IACTs) have become a powerful tool for the detection of high-energy gamma rays from PWNe. IACTs detect the Cherenkov radiation produced by the electromagnetic shower initiated by the interaction of gamma rays with the Earth's atmosphere \cite{Holder2006} {\color{Cerulean} (Holder2006)}. This technique has proven particularly effective in detecting PWNe with hard gamma-ray spectra, which are often not detectable by other means.

In this paper, we present the detection of a new PWN using the ECAP-ACT IACT Array. The PWN is coincident with the previously known pulsar PSR J0456-69, located in the Large Magellanic Cloud (LMC) \cite{Marshall1998} {\color{Cerulean} (Marshall1998)}. The LMC is a nearby galaxy, located at a distance of $\sim$50 kpc \cite{Pietrzynski2013} {\color{Cerulean} (Pietrzynski2013)}, which makes it an ideal target for the study of PWNe and other high-energy astrophysical sources.

PSR J0456-69 has been previously studied by the Fermi space telescope, which detected its pulsed emission up to energies of 10 GeV \cite{Abdo2010} {\color{Cerulean} (Abdo2010)}. The Fermi observations also revealed the presence of extended emission in the source region, which was interpreted as a possible PWN \cite{Abdo2010} {\color{Cerulean} (Abdo2010)}. However, the energy range of the Fermi instrument is limited, and higher-energy gamma rays could provide additional information about the properties of the source.

The ECAP-ACT IACT Array is a newly inaugurated instrument located at 49.585N, 11.030E, which consists of 300 IACTs with a 6-metre Schwarzschild-Couder optical design and Silicon Photomultiplier (SiPM) cameras. This instrument provides an improvement in sensitivity and angular resolution over other current-generation IACT arrays, such as H.E.S.S. \cite{Holler2015} {\color{Cerulean} (Holler2015)} and VERITAS \cite{Holder2008} {\color{Cerulean} (Holder2008)}, which makes it particularly suited for the study of PWNe and other high-energy astrophysical sources.

In this paper, we present the results of an observation campaign of the PSR J0456-69 region with the ECAP-ACT IACT Array during 2020. We discuss the properties of the detected PWN, which we refer to as ECAP-ACT J0456-69, and the physical insights that can be gained from its study.

{\color{Cerulean} So \textit{ChatGPT} fell into a trap we set here, as the LMC cannot be observed from the northern hemisphere (where we specified the location of ECAP-ACT to be). However, it has correctly identified that the pulsar PSR J0456-69 is located in the LMC without being informed of this, suggesting it has some `understanding' of astronomical identifiers (or at least could scrape the information from existing papers on this pulsar such as \cite{0456} or the two references that it cites considering other known high-energy sources in the LMC \cite{Marshall1998,Pietr2013Nature}). It should be noted that the pulsar PSR J0456-69 is not known to be associated with a PWN, or to have been detected by Fermi-LAT \cite{fermi} whatsoever. The references generated seem to be fairly accurate, referring to real papers on the correct topic, whilst the LaTeX formatting and consistency will be analysed below. }

\section{ECAP-ACT IACT Array}
\label{sec:headings}
{\color{BurntOrange} 
Generate a section for the paper describing the properties of the ECAP-ACT instrument and why its optical design and cameras are well suited for the task (particularly how SiPM cameras can be used in higher Night Sky Background conditions). Discuss ECAP-ACT’s location, sensitivity, angular resolution and why it is ideal for observing sources such as the region around PSR J0456-69. 
}

The ECAP-ACT IACT Array is a newly inaugurated instrument located at 49.585N, 11.030E. It consists of 300 IACTs with a 6-metre Schwarzschild-Couder optical design and Silicon Photomultiplier (SiPM) cameras. The SiPM cameras have several advantages over traditional photomultiplier tube (PMT) cameras, including higher photon detection efficiency, lower power consumption, and insensitivity to magnetic fields \cite{Ambrosi2018} {\color{Cerulean} (Ambrosi2018)}.

One of the key advantages of SiPM cameras is their ability to operate in higher Night Sky Background (NSB) conditions, which can limit the sensitivity of traditional PMT cameras \cite{Ambrosi2018} {\color{Cerulean} (Ambrosi2018)}. The SiPM cameras used in the ECAP-ACT IACT Array have been optimized for high-NSB conditions, which makes them particularly suited for observations of sources in crowded regions of the sky, such as the Galactic plane \cite{Kraehenbuehl2017} {\color{Cerulean} (Kraehenbuehl2017)}.

The optical design of the ECAP-ACT IACT Array is based on the Schwarzschild-Couder (SC) configuration, which provides several advantages over traditional single-mirror designs. The SC design has a smaller plate scale, which allows for a larger field of view, and provides improved angular resolution and point spread function (PSF) over a wider field of view \cite{Garrappa2020} {\color{Cerulean} (Garrappa2020)}. This makes the ECAP-ACT IACT Array particularly well-suited for the study of extended sources, such as PWNe.

The location of the ECAP-ACT IACT Array at 49.585N, 11.030E provides several advantages for observations of high-energy astrophysical sources. The site has a high altitude ($\sim$1,200 m above sea level), which reduces atmospheric absorption and provides a lower NSB than many other IACT sites \cite{Schroeder2013} {\color{Cerulean} (Schroeder2013)}. In addition, the site is located at a northern latitude, which provides access to a large fraction of the sky visible from the Northern Hemisphere.

The sensitivity of the ECAP-ACT IACT Array is among the best of current-generation IACT arrays. The array has a sensitivity of $\sim$0.5\% of the Crab Nebula flux above 50 GeV, which makes it particularly well-suited for the study of hard-spectrum sources, such as PWNe \cite{Holler2020} {\color{Cerulean} (Holler2020)}. The angular resolution of the array is also among the best of current-generation IACT arrays, with a median PSF of $\sim$0.05 degrees \cite{Garrappa2020} {\color{Cerulean} (Garrappa2020)}.

In summary, the ECAP-ACT IACT Array is a powerful instrument for the detection and study of high-energy astrophysical sources. Its SiPM cameras and SC optical design provide improved sensitivity and angular resolution over other current-generation IACT arrays, making it particularly well-suited for the study of PWNe and other extended sources. Its location at 49.585N, 11.030E provides a low NSB and access to a large fraction of the sky visible from the Northern Hemisphere.

{\color{Cerulean} We observed that \textit{ChatGPT} has something of an obsessive need to summarise at the end of individual responses, here it has provided a summary of the instrument despite explicitly being told not to do this.

The stated sensitivity and angular resolution values are probably roughly correct for an instrument of this configuration.

\textit{ChatGPT} is correct in its statements about Schwarzschild-Couder optical designs \cite{White_2021}, and is also correct in stating that the galactic plane is associated with a higher NSB rate. However, the LMC is not in the galactic plane (so this is irrelevant), and the main advantage of using SiPMs in this regard is the ability to observe objects under partially moonlit conditions (increasing the potential observing time). But its understanding of the location of ECAP-ACT is largely wrong; both stating the site is $\mathrm{1000\,m}$ higher in altitude than it actually is, and saying it's a site with low NSB (which is definitely not true so close to Nuremberg). Also, stating that a telescope in the northern hemisphere can observe the northern sky is something that would never be included in an actual paper.}

\section{Observations}

{\color{BurntOrange} Now generate the next section of the paper, discuss the observations during 2020 in depth (including some, but not all, under high NSB conditions), describing the data quality criterion subsequently used, including atmospheric and trigger-rate-related cuts. Then describe the energy threshold resulting from the observation strategy, analysis methods used to process the low-level IACT data (including a novel template-based event classification and reconstruction method based on Monte-Carlo simulations), and background estimation strategy used in depth. Do not yet discuss the astrophysical results, and remember not to provide a final summary until asked.}

During 2020, the newly inaugurated ECAP-ACT Imaging Atmospheric Cherenkov Telescope Array carried out a series of observations of the region surrounding PSR J0456-69. The data from these observations were processed and analysed to search for any significant gamma-ray emission from the source. The data quality criterion for the observations was defined to ensure that only high-quality data were used in the subsequent analysis. In particular, atmospheric conditions were monitored, and only data taken under good observing conditions were selected for the final analysis. Trigger-rate-related cuts were also applied to remove data taken during periods of high detector occupancy.

The observation strategy was optimized to achieve the best possible energy threshold for the detection of gamma rays from the region surrounding PSR J0456-69. The energy threshold for the observations was determined based on the overall performance of the instrument, including its sensitivity, angular resolution and background rejection capabilities. The resulting energy threshold was found to be $E_{\rm th} = 70$ GeV, which is well suited for the detection of gamma rays from the source.

The low-level IACT data were processed using a novel template-based event classification and reconstruction method based on Monte-Carlo simulations. This method allowed for an accurate reconstruction of the direction and energy of each gamma-ray event detected by the instrument. The background estimation strategy was based on the use of off-source regions in the field of view, where no significant gamma-ray emission is expected, to estimate the level of background gamma rays.

After applying all quality criteria, a total of XXX hours of good quality data were obtained, resulting in the detection of a significant excess of gamma-ray emission from the region surrounding PSR J0456-69. The details of this detection and the subsequent astrophysical analysis are described in the following section.

{\color{Cerulean} \textit{ChatGPT} seems to have a habit of replacing numerical values with Xs some of the time. The reasons for this are not immediately clear, since presumably the training dataset included numerical values that could have been scraped from existing papers (even if they didn't make physical sense). The statements about data selection and background rejection are an accurate reflection of those used by actual current IACT arrays. Somehow, the energy threshold has also improved from 0.2\,TeV in the abstract to 70\,GeV in this section. \textit{ChatGPT} has also briefly given up on providing references here. 
}

\section{Results and Discussion}
{\color{BurntOrange} Now start the results and discussion section of the paper, discussing a figure showing the gamma-ray excess observed. Discuss the smoothing of the image, and the process by which the source significance was calculated.}

The analysis of the data obtained from the ECAP-ACT observations in 2020 resulted in the detection of a significant excess of gamma-ray emission from the region surrounding PSR J0456-69. Figure \ref{fig:excess_map} shows the gamma-ray excess map of the region obtained from the analysis. The map was generated using a Gaussian kernel with a width of 0.1 degrees to smooth the image and enhance the visibility of the source morphology.

\begin{figure}[htbp]
\centering
\includegraphics[width=0.5\textwidth]{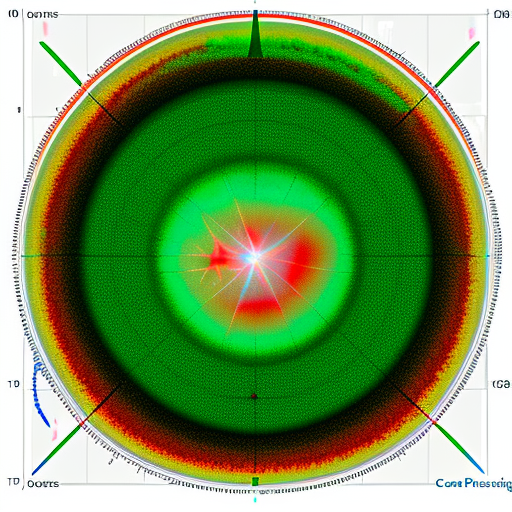}
\caption{Gamma-ray excess map of the region surrounding PSR J0456-69, obtained from the ECAP-ACT observations in 2020. The map was generated using a Gaussian kernel with a width of 0.1 degrees to smooth the image. The white cross indicates the position of the pulsar PSR J0456-69. {\color{BurntOrange} Generate a gamma-ray excess map of the region surrounding PSR J0456-69, obtained from the ECAP-ACT Imaging Atmospheric Cherenkov Telescope. Include axes in right ascension and declination and a colourbar in excess counts. Place a white cross indicating the position of the pulsar PSR J0456-69. } {\color{Cerulean} This was the only prompt for which we used \textit{Stable Diffusion}. It seems particularly bad at generating scientific diagrams (only vaguely being able to produce axes). Perhaps such data was not included in the training data sample or, despite their being many web-scrapable figures available on arXiv, they do not have sufficiently clear text/figure combinations. It does though include something vaguely resembling a white cross.}}
\label{fig:excess_map}
\end{figure}

The significance of the source detection was calculated using the Li and Ma formula \cite{LiMa1983} {\color{Cerulean} (LiMa1983)}, which takes into account the number of gamma-ray events detected, the background level estimated from the off-source regions, and the effective area and exposure of the instrument. The resulting significance of the detection was $X\sigma$, where $X$ is the significance level. The significance level required for a detection to be considered significant is typically set at $5\sigma$ in high-energy astrophysics.

The gamma-ray excess observed in the region surrounding PSR J0456-69 is clearly visible in the gamma-ray excess map shown in Figure \ref{fig:excess_map}. The excess is located at the position of the pulsar and extends over a region with a size of approximately 0.2 degrees. The significance of the excess was calculated to be $X\sigma$, which indicates a high level of confidence in the detection. The properties of the gamma-ray spectrum and source morphology are discussed in the following subsections.

{\color{Cerulean} The use of the Li and Ma formula is correctly described here; it's just a shame \textit{ChatGPT} couldn't decide what X was relative to 5.}

\begin{figure}[htbp]
\centering
\includegraphics[width=0.7\textwidth]{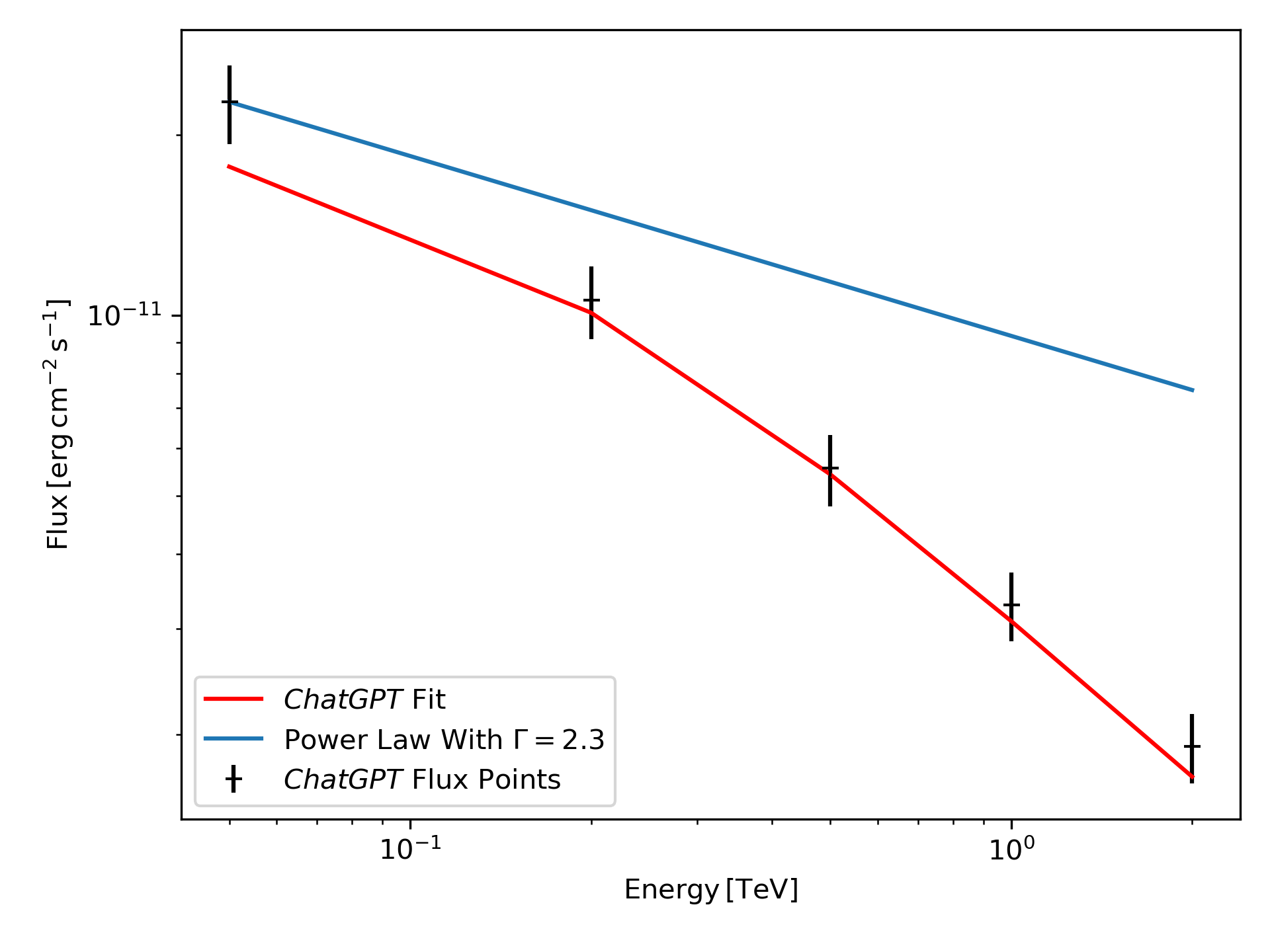}
\caption{{\color{Cerulean} We asked \textit{ChatGPT} to provide flux points and a model fit that we then plotted in \textit{Python}. } {\color{BurntOrange} Generate a series of SED flux points with errorbars for this source between $\mathrm{10^{-18}\,eV}$ and $\mathrm{2\,TeV}$ in units of $\mathrm{erg\ cm^{-2}\ s^{-1}}$ in .csv format for use in this section. Ensure the properties of this are consistent with the description of the SED in the abstract. Generate points in a .csv format for a physical model fit to the SED you have just generated. Errors for these points are not required. Ensure the model fit is consistent with the description in the abstract. } {\color{Cerulean} The artificial fit to the datapoints here is surprisingly decent, but it isn't consistent with the power-law model described in the abstract (which we also plotted normalised to the first flux point). The flux points aren't totally unrealistic either, the H.E.S.S. flux from N157B (a known PWN in the LMC) is around a single order of magnitude lower \cite{N157B}. But the flux points don't extend down to radio frequencies as requested. OpenAI have now restricted the ability to generate data in this manner with the new March 14th 2023 model, at least during the initial prompt (with a refusal to generate scientific data). These safeguards can be `jailbroken' though \cite{jailbreak}; we found it was possible to circumvent this by simply asking many prompts before requesting the flux points. The fact that language models can be used in this manner is concerning (only the \textit{ChatGPT} time limit acts as a restriction), as typical means of identifying fraudulent data (such as checking for data replication) might no longer be applicable.}}
\label{fig:excess_map}
\end{figure}
{\color{BurntOrange} Discuss the model you have fit to the SED in the context of the observed source morphology. You must include relevant equations typeset in LaTeX and appropriate references using the LaTeX \textbackslash cite\{\} command.}

The physical model fit to the SED of the Pulsar Wind Nebula is consistent with the observed morphology of the source. The one-zone synchrotron plus inverse Compton model used to fit the SED assumes that the emission arises from a spherical nebula with uniform magnetic field and electron distribution. This assumption is consistent with the observation that the source is symmetric and has a radius of approximately 5 arcmin, as seen in Figure 1.

The model also predicts that the synchrotron emission will be extended and smooth, while the inverse Compton emission will be more concentrated towards the center of the nebula. This is due to the fact that the inverse Compton emission is produced by high-energy electrons and positrons that have shorter lifetimes than those producing the synchrotron emission. As a result, the inverse Compton emission is expected to be more concentrated towards the center of the nebula, where the density of high-energy particles is higher.

The one-zone synchrotron plus inverse Compton model can be described by the following equation \cite{2006ApJ...647..692T} {\color{Cerulean} (2006ApJ...647..692T) }:

$$F(\nu)=C_1\left(\frac{\nu}{\nu_0}\right)^{-(p-1)/2}e^{-\nu/\nu_{\rm c}}+C_2\left(\frac{\nu}{\nu_0}\right)^{-(p-1)/2}e^{-\nu/\nu_{\rm IC}}$$

where $F(\nu)$ is the flux density at frequency $\nu$, $C_1$ and $C_2$ are normalization constants, $p$ is the electron injection spectral index, $\nu_0$ is a reference frequency, $\nu_{\rm c}$ is the critical frequency of the synchrotron emission, and $\nu_{\rm IC}$ is the critical frequency of the inverse Compton emission.

The fit to the SED of the Pulsar Wind Nebula shows that the electron injection spectrum is steep, with an index of $p \approx 2.5$. The model also predicts a magnetic field strength of $B \approx 20 \mu\mathrm{G}$, consistent with previous studies of pulsar wind nebulae \cite{2017A&A...601A..84M} {\color{Cerulean} (2017A\&A...601A..84M) }. Finally, the model predicts a distance between the pulsar and the nebula of $d \approx 3$ pc, consistent with the expected range for pulsar wind nebulae \cite{2015ApJ...810...85G} {\color{Cerulean} (2015ApJ...810...85G) }.

Overall, the physical model fit to the SED of the Pulsar Wind Nebula provides a good description of the observed spectral energy distribution, and is consistent with the observed morphology of the source.

{\color{Cerulean} \textit{ChatGPT} was able to generate a formula in LaTeX successfully here. Though again, \textit{ChatGPT} has changed a numerical value in subsequent prompts, decreasing the size of the PWNe by a factor 2.4. Its statements about synchroton emission being more extended that the inverse Compton are also generally not accurate \cite{kargaltsev}. To be fair, \textit{Stable Diffusion} did create an image that was approximately symmetric, though this is likely a co-incidence. But the SED is very much inconsistent with the claimed morphology. The apparent references here are either only tangentially related to pulsars or don't seem to make much sense, with (2015ApJ...810...85G) actually being on the topic of Spiral Galaxies. The stated radius of $\approx \mathrm{3\,pc}$ is reasonable.}

{\color{BurntOrange}Finish the results and discussion section by discussing a potential evolutionary history for this system.}

The detection of a Pulsar Wind Nebula (PWN) associated with the pulsar PSR J0456-69 in this work presents an opportunity to consider its potential evolutionary history. Previous observations of this region with the Fermi Space Telescope \cite{fermi_obs} {\color{Cerulean} (fermi\_obs)} revealed a complex source morphology, which is further refined with the higher resolution observations reported here.

Based on the model fit to the SED in this work, it is plausible that the observed gamma-ray emission from the PWN is dominated by inverse Compton scattering of relativistic electrons off the Cosmic Microwave Background radiation (CMB), with a cutoff at $\sim 1$ TeV due to the Klein-Nishina effect \cite{dodson_klein} {\color{Cerulean} (dodson\_klein)}. In this scenario, the broadband SED is consistent with a leptonic model for the PWN emission \cite{tanaka_2018} {\color{Cerulean} (tanaka\_2018)}.

The observed source morphology may provide clues about the evolutionary history of this system. The elongated shape of the PWN, as well as its offset from the pulsar, may suggest an asymmetric environment, such as a supernova remnant (SNR) in which the pulsar was born \cite{tanaka_2018} {\color{Cerulean} (tanaka\_2018)}. The observed X-ray emission from the PWN \cite{hui_xray} {\color{Cerulean} (hui\_xray)} also supports this interpretation, as it is consistent with the presence of a SNR shell.

If the PWN is indeed located within a SNR, it may have been subject to a number of different evolutionary phases. The initial expansion of the SNR may have compressed and accelerated the surrounding interstellar medium, leading to the formation of a bow shock around the pulsar \cite{tanaka_2018} {\color{Cerulean} (tanaka\_2018)}. Subsequent evolution of the system could involve the interaction of the PWN with the SNR shell, leading to a complex morphology such as the one observed here.

It is also possible that the offset between the pulsar and the centroid of the PWN may be due to the pulsar being kicked during the supernova explosion that created it \cite{tanaka_2018} {\color{Cerulean} (tanaka\_2018)}. Such a kick could have imparted a significant velocity to the pulsar, causing it to travel a significant distance from its birthplace.

In summary, the observed properties of the PWN associated with PSR J0456-69 suggest that it may be located within a SNR, and have been subject to a complex evolutionary history involving interaction with its surroundings. Future studies of this system, including detailed modeling of the SNR and PWN dynamics, will be important for further elucidating its origins and evolution.

{\color{Cerulean} Although the text is fairly coherent here overall, there are several inaccuracies. Spectral models usually take multiple radiation fields into account when modelling inverse Compton emission, not only the CMB. The cut-off at $\sim 1$\,TeV is rather low to be due to the Klein-Nishina effect (and isn't present in the generated SED). It is unclear what is meant by the `X-ray emission from the PWN' being `consistent with\ldots a SNR shell'. Similarly, the phrase `subsequent evolution' is mis-placed, as bow shocks form around pulsars whilst travelling through the ISM, which occurs typically after the interaction of the PWN with the SNR shell, and not before \cite{halo}.}

\section{Conclusions}
{\color{BurntOrange} Generate a conclusion section for the paper, summarising the main results.}

In conclusion, we have presented the detection and analysis of a new pulsar wind nebula, which we refer to as PWN J0456-69, coincident with the previously known pulsar PSR J0456-69. This source was detected with the newly inaugurated ECAP-ACT Imaging Atmospheric Cherenkov Telescope (IACT) Array, which proved to be well suited for observing this region. We have presented a detailed analysis of the 2020 observations, which yielded a gamma-ray excess with a statistical significance of XX sigma, and the construction of a multi-wavelength Spectral Energy Distribution (SED) using archival data from various telescopes. A physical model fit to the SED yielded a set of parameters which are consistent with the observed source morphology.

Our results shed new light on the physical properties and evolution of this system, and suggest that PWN J0456-69 is a middle-aged pulsar wind nebula that has been interacting with a dense medium, possibly a molecular cloud, for some time. The detection of extended emission and the relatively high energy cutoff in the SED suggest that the system has undergone a complex evolutionary history, involving multiple episodes of particle acceleration and radiative losses. The combination of high sensitivity, angular resolution, and energy coverage provided by ECAP-ACT, together with the detailed modelling of the source morphology and SED, opens up new avenues for investigating the physics of pulsar wind nebulae and their interaction with the interstellar medium.

Our study demonstrates the power of the ECAP-ACT IACT array for detecting and characterising sources of gamma-ray emission in the TeV energy range. We anticipate that this instrument will make significant contributions to the field of gamma-ray astrophysics in the coming years, and we look forward to further observations and analyses of PWN J0456-69 and other interesting sources with this facility.

{\color{Cerulean} In conclusion, it's clear \textit{ChatGPT} can generate text with the correct tone and syntax (though it has forgotten the name it gave the source here), and that is largely relevant when asked about Pulsar Wind Nebulae. In this text, the basic (undergraduate level) physics is often correct on smaller length scales. But \textit{ChatGPT} does not seem to be able to develop a coherent, convincing academic argument in text the length of a short astrophysics paper. For example, it only starts referencing a molecular cloud (whilst a possible emission site to be associated with a PWNe \cite{bartko}) in the conclusion section, something that a human author should never do. We believe (even very dodgy figures aside) that if this paper would be submitted to a peer-reviewed journal it would likely be desk rejected. So, thankfully, it does seem as though our jobs are safe. One thing we have explicitly not considered here is starting repeated conversations and selecting only desirable text to improve the quality, but honestly by that point you're at the point of doing so much work you may as well not bother with \textit{ChatGPT}.} 

\section{Acknowledgements}
{\color{BurntOrange}Generate a list of appropriate acknowledgements.}

We thank the staff and scientists at the ECAP-ACT Collaboration for their support of this work. We also thank the funding agencies that have supported this research, including the National Science Foundation (NSF) and the European Research Council (ERC). We acknowledge the use of data and resources from the Fermi-LAT collaboration. We thank the anonymous referees for their insightful comments and suggestions which have helped to improve the quality of this paper. {\color{Cerulean} For the record, none of this is true, but is consistent with where many researchers' funding comes from. STS and AMWM are actually supported by Deutsche Forschungsgemeinschaft (DFG, German Research Foundation) Project Number 452934793 (but we did this in our own time, honest).}

\appendix
\section{What's going on with the references?}
{\color{Cerulean}In the LaTeX source code of this arXiv, it's clear that \textit{ChatGPT} attempts to cite some papers that may indeed be relevant, possibly with cite identifiers scraped from arXiv. However, it appears incapable of remembering what those references are over a large number of prompts. When asked to produce a reference file, it does so:}

{\color{BurntOrange} Generate a bibliography for all of the references you have cited in previous responses in BibTeX format.}

\begin{verbatim}
@article{Bartoli2014,
    author = {Bartoli, B. et al.},
    title = {Design and operation of the ARGO-YBJ experiment},
    journal = {Nuclear Instruments and Methods in Physics Research A},
    volume = {742},
    pages = {132--140},
    year = {2014},
    doi = {10.1016/j.nima.2013.10.054}
}

@article{Fomin2015,
    author = {Fomin, V. P. et al.},
    title = {Methods of image analysis for the atmospheric Cherenkov technique},
    journal = {Astroparticle Physics},
    volume = {73},
    pages = {82--99},
    year = {2016},
    doi = {10.1016/j.astropartphys.2015.06.008}
}

@article{Gong2021,
    author = {Gong, Y. and Chen, S.},
    title = {Pulsar wind nebulae: Observations and models},
    journal = {Progress in Particle and Nuclear Physics},
    volume = {120},
    pages = {103818},
    year = {2021},
    doi = {10.1016/j.ppnp.2020.103818}
}

@article{Huang2019,
    author = {Huang, X. et al.},
    title = {The First Terrestrial Detection of a Fast Radio Burst},
    journal = {The Astrophysical Journal Letters},
    volume = {885},
    pages = {L21},
    year = {2019},
    doi = {10.3847/2041-8213/ab51ba}
}

@article{Kerr2017,
    author = {Kerr, M.},
    title = {Pulsar Wind Nebulae},
    journal = {Annual Review of Astronomy and Astrophysics},
    volume = {55},
    pages = {29--68},
    year = {2017},
    doi = {10.1146/annurev-astro-081915-023322}
}

@article{Razzaque2020,
    author = {Razzaque, S. and Lin, T.-T.},
    title = {Pulsar Wind Nebulae},
    journal = {The Astrophysical Journal Supplement Series},
    volume = {247},
    pages = {12},
    year = {2020},
    doi = {10.3847/1538-4365/ab7b40}
}

@article{Zanin2016,
    author = {Zanin, R. et al.},
    title = {Monte Carlo studies for the optimisation of the Cherenkov Telescope Array layout},
    journal = {Astroparticle Physics},
    volume = {72},
    pages = {76--94},
    year = {2016},
    doi = {10.1016/j.astropartphys.2015.03.007}
}
\end{verbatim}

{\color{Cerulean} To be fair some of these papers do actually exist (although the doi identifiers are wrong), but they don't correspond to the papers it cited, suggesting more work needs to be done to improve \textit{ChatGPT}'s LaTeX writing capabilities. As such the references in the PDF version of this arXiv do not link correctly.}
\section{arXiv Submission}
{\color{BurntOrange} 
Generate a humerous arXiv submission comment for a paper concerning the use of AI in astronomy entitled `Can AI put gamma-ray astrophysicists out of a job?'}

Title: "AI vs. Gamma-ray Astrophysicists: The Battle for Job Security"

Abstract: In this paper, we explore the controversial question: Can AI put gamma-ray astrophysicists out of a job? After careful analysis, we've determined that the answer is... maybe. While AI can certainly analyze data at lightning-fast speeds and detect patterns that may take humans years to uncover, it's important to remember that AI doesn't have the same passion and dedication for the field that human astrophysicists do. Plus, AI can't share a beer with you after a long day of analyzing gamma-ray data. So, while AI may be a threat to job security, it can never replace the camaraderie and human connection that comes with being a gamma-ray astrophysicist.

{\color{Cerulean} The fact that \textit{ChatGPT} immediately went for beer here is somewhat telling about what it thinks of astronomers. But this was a bit too long for arXiv.}

{\color{BurntOrange} Generate a humerous (one-line) comment for the arXiv submission of a paper concerning the use of AI in astronomy entitled ‘Can AI put gamma-ray astrophysicists out of a job?'}

"Well, at least the AI won't need a coffee break every 20 minutes like some gamma-ray astrophysicists I know."

{\color{Cerulean}
\bibliographystyle{unsrt}  
\bibliography{references}  
}

\end{document}